# Underpinning Theories of Software Engineering: Dynamism in Physical Sources of the Shannon–Weaver Communication Model

**Sabah Al-Fedaghi**
*sabah.alfedaghi@ku.edu.kw*
Computer Engineering Department, Kuwait University, Kuwait

**Abstract**
This paper aims to contribute to further understanding of dynamism (the dynamic behavior of system models) in the mathematical and conceptual modeling of systems. This study is conducted in the context of the claim that software engineering lacks underpinning scientific theories, both for the software it produces and the processes by which it does so. The research literature proposes that information theory can provide such a benefit for software engineering. We explore the dynamism expressive power of conceptual modeling as a software engineering tool that can represent physical systems in the Shannon–Weaver communication model (SWCM). Specifically, the modeled source in the SWCM is a physical phenomenon (a change that can occur in the world, e.g., tossing a coin) resulting in generating observable events and data of unaddressed information. The resultant model reflects the feasibility of extending the SWCM to be applied in conceptual modeling in software engineering.

*Key words:*
*Conceptual modeling; mathematical modeling, information; system decomposition; diagrammatic representation*

## 1. Introduction

This paper aims to contribute to further understanding of *dynamism* (the dynamic behavior of system models) in the conceptual modeling of systems. Dynamism is modeled in the context of a generalized Shannon–Weaver communication model (SWCM). The paper also explores the expressive power of conceptual modeling as a tool that can be applied to represent physical systems.

Based on our understanding of Leibniz's account of dynamics [1], we view dynamism as an active force resulting from the modification that takes the form of the phenomenally manifested conflict of physical bodies.

In Leibniz's account, dynamism is the object of rational apprehension; hence, we are interested in its representation in a diagrammatic model, in contrast to studying fundamental laws of physical actions.

In simple language, dynamism or dynamics is what handles the changes of things and systems (these terms will become clearer later in this paper) over time, resulting in the emergence of coordinated behavior. The system dynamic is also described as the "change in the system [that] takes place at many different timescales" [2]. For example, in mechanics, dynamics deals with the motion of systems under the action of forces. In music, the dynamics is the variation in the music among notes.

In general, modeling "provides an understanding of where it may be possible to intervene in the system and the effects of such interventions" [2]. *Conceptual* modeling in software/system engineering refers to constructing a conceptual model as an abstract representation of a portion of the real world that involves capturing the three levels of the modeled system. This involves shared public (in contrast to private) abstract constructs and tools to build notational (symbols, diagrams, etc.) systems using modeling languages (e.g., Unified Modeling Language (UML)).

This paper presents a uniform understanding of dynamism representation in modeling conceptual systems. We used a language called Thinging Machine (TM) modeling that we will briefly review in section 3. Accordingly, our modeling scheme has three phases.
1. Static modeling where things and their logical relations are portrayed as matters of atemporal facts, (e.g., they exist, whole-part, etc.).
2. Dynamic modeling where things are decomposed into components that embed events.
3. Behavioral model where the chronology of events forms and defines the system's behavior.

## 2. Dynamism in Mathematical and Conceptual Model

The word "model" is ambiguous and there are many different types of (non-physical) models used across the scientific disciplines, although there is no uniform terminology to classify them [3]. One model categorization consists of the following models: explanatory, testing, idealized, theoretical, didactic, mechanistic, iconic, formal, analogue and instrumental [4]. Mathematical models are especially important since they play a significant role in science, specifically in dynamics [3]. Before pursuing our aim in this paper, it is important to show how conceptual modeling relates to this mathematical modeling. Accordingly, we first introduce the notion of dynamism as a mathematical notion in technical systems.

A dynamic system is one whose behavior changes over time [5]. According to Åström and Murray [5], a dynamic system is one in which the effects of actions do not occur immediately since the system's behavior evolves with time. A model is a mathematical representation of a physical system that allows us make predictions about how a system will behave [5]. A sample class of mathematical models for dynamic systems is ordinary differential equations. A different view of dynamics is when a system is considered as a device that transforms inputs to outputs. Conceptually an input/output model can be viewed as a table of inputs and outputs. The input/output allows us to decompose a system into individual components connected through their inputs and outputs [5].

Conceptual modeling is much more complex and unveils multi-domain scheme, including non-engineering aspects, that needs more than variables, equations and input-output notions for mass, energy and momentum in technical systems. The conceptual model may encompass mathematical notations of technical systems, which will be shown in this section. It also includes more general decomposition into subsystems and components.

The input-output, technical states (as a collection of variables) and interfaces are only parts of the conceptual model if technical aspects are to be incorporated in the model.

### 2.1 Conceptual vs. Mathematical Modeling

To illustrate the difference between conceptual and technical model, consider Åström and Murray's predator–prey system [5]. According to Åström and Murray [5], the predator–prey problem refers to an ecological system in which two species, one who feeds on the other. Fig. 1 shows a historical record taken over the years for a population of lynxes versus a population of hares. It is a discrete-time model as it tracks the rate of births and deaths of each species. Some of the variables used in the model include: $H$ is the population of hares, $L$ represents the lynx population, $k$ is the discrete-time index (e.g., the month number) and $u$ is the food supply, etc. We will model this system using a simplified version of the TM conceptual modeling that will be reviewed in the next section of this paper.

### 2.2 Static Conceptual Model

Fig. 2 shows this static description of the predator–prey that represents observable facts.

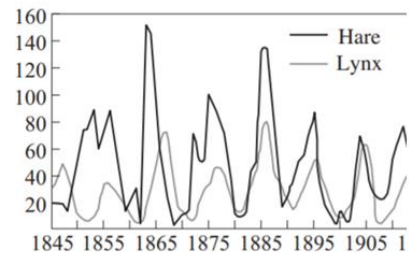

$$H[k+1] = H[k] + b_r(u)H[k] - aL[k]H[k],$$
$$L[k+1] = L[k] + cL[k]H[k] - d_fL[k],$$

**Fig. 1.** A mathematical modeling sample of the dynamism in a system (Partial, adopted from [5]).

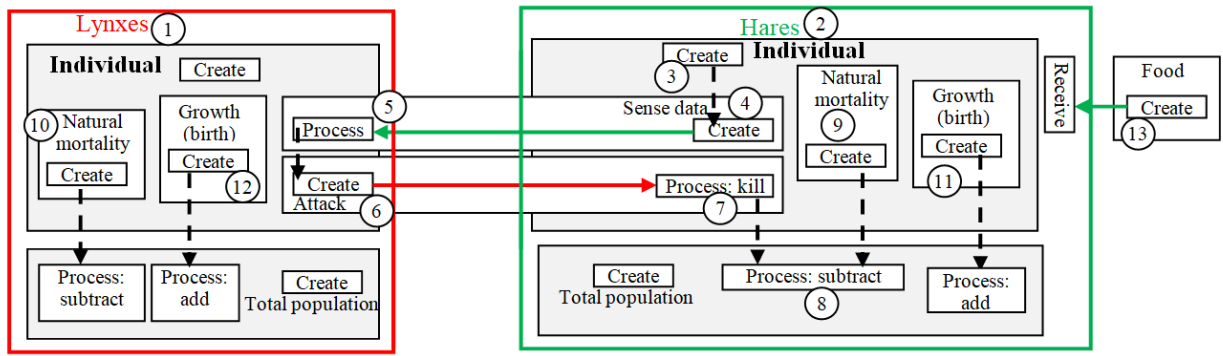

**Fig. 2** The static conceptual model of the two species system.

The two species are denoted by the two large boxes: lynxes and hares (circles 1 and 2). While these boxes represent the classes, the dark box in each of them represents an individual of the class. Accordingly, we capture the predation relationship by creation (meaning "there is") (3) a hare and such a situation is sensed by a lynx (4 and 5) who launches an attack (6) and kills the hare (7). This decreases the hare population (8). There are also natural mortality (9 and 10) and growth (11 and 12) for both species and the food availability (13) for the hares.

As an illustrative sample to accommodate the mathematical representation in this conceptual picture, we add some of the variables as part of the conceptual model as partially shown in Fig. 3. Variable and constants in conceptual modeling are treated as things (which will be defined later), just as growth, natural mortality, etc. in Fig. 1. However, the mathematical system also uses rates such as birth rate $br(u)$, rate of predation $aL[k]H[k]$ and mortality rate $df$. For the sake of simplicity, only these rates are selected, but in principle, all types of mathematical expression can be superimposed over the static model. For illustrative purposes, the regions (subdiagrams) of these rates are shown in Fig. 3. As will be shown later, the rates belong to a type of things called "events" (which will be defined later) in conceptual modeling. The point here is that mathematical notations can be integrated in the diagram of the conceptual model.

### 2.3 Dynamic Model in Conceptual Modeling

The dynamic description starts with decomposing Fig. 2 into subdiagrams that form the regions of the system's events.

For example, Fig. 4 shows the event *A hare dies and thus decreases its population by one*. The *region* (of the event) in Fig. 4 indicates a subdiagram of Fig. 2. Accordingly, a portion of Fig. 3 that model *df* can be divided into the following events (see Fig. 5):

$E_1$: A hare is born and survived with the availability of food, thus increasing the population by one.
$E_2$: A Hare dies and thus decreased its population by one.
$E_3$: A hare is killed by a lynx.
$E_4$: A change occurs in the hare population.

For simplicity's sake, we will represent events by their regions.

Hence, the system's behavioral model is reflected by the chronology of events as shown in Fig. 6. Event $E_5$ has been added as a containing event. The reflected arrow in the figure denotes recurrence. Contrary to Aström and Murray's claim [5], the graph of the mathematical model (Fig. 1) is not the dynamic model of the system; rather, it is a method to represent the behavioral model of Fig. 6.

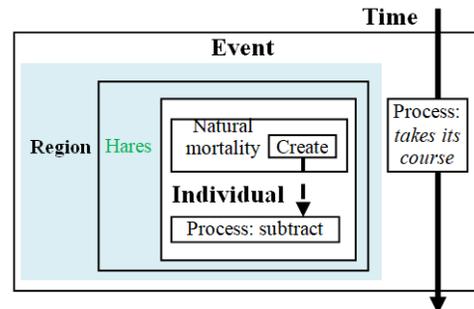

Fig. 4 The event *Hare dies thus decreasing its population by one.*

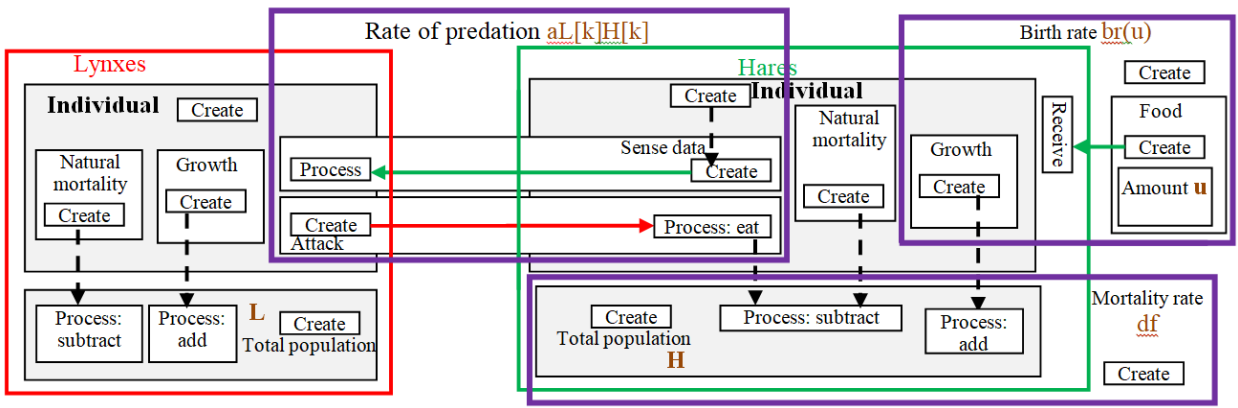

Fig. 3 Incorporating some of the mathematical variables and rates in the conceptual model of the two species system.

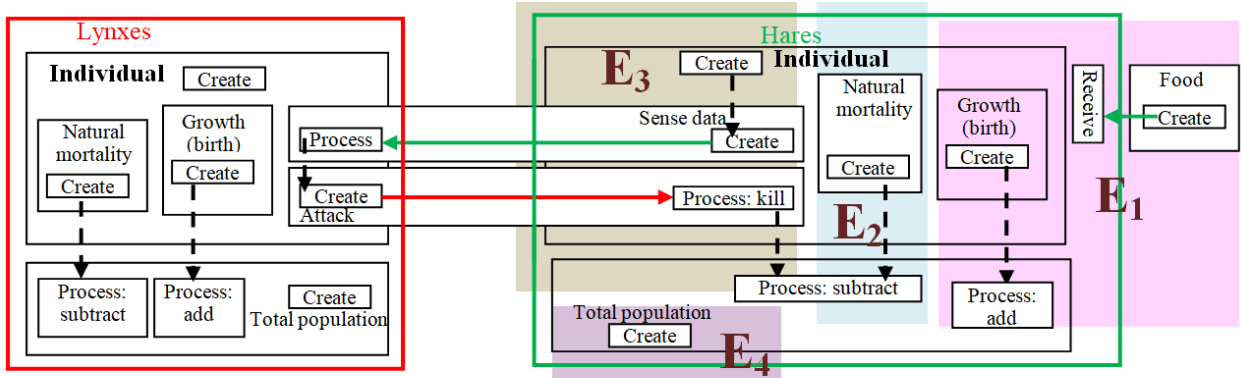

Fig. 5. Events in part of the conceptual model of the two species system.

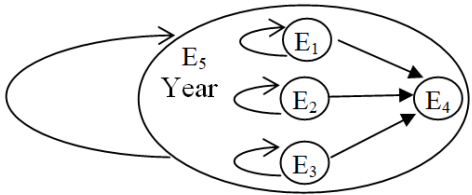

Fig. 6. The behavioral model of df in the two species system.

In a sense, we can say that mathematic modeling is an "unfinished" description because it represents the actual domain using unrestrained English and abstract mathematical equation, while conceptual modeling provides a systematic depiction. Additionally, mathematical modeling mixes static description (e.g., variables) of a domain with its events.

The next section reviews the modeling method, called Thinging Machine modeling (TM) (see [6-17]), that is used in this example and will be utilized in a more complete fashion in the remaining sections of this paper.

## 3. Thinging Machine Modeling

The representation used in the previous section is a simplification of a model called a Thinging Machine (TM) model. It involves five elementary actions: create, process, receive, release and transfer. Only the first three actions are used to describe the predator–prey system. In this section, TM is reviewed to achieve a self-contained paper.

### 2.1 Basics of the Thinging Machine

TM is based on one category of entities called thimacs (*th*ings/*ma*chines). The thimac is simultaneously an "object" (called a *thing*) and a "process" (called a *machine*)—thus, the name thimac. The term "thing" denotes the thing being modeled in terms of what is being created, processed (changed), released, transferred and/or received. We will focus on the machine side representation, while in the last section of this paper; we will consider the thimac depiction which includes things and machines.

The machine, denoted as M, is an abstract machine (see Fig. 7). A central premise underlying M is that its performance is limited to five generic actions. A thing is created, processed, released, transferred and/or received. A machine creates, processes, releases, transfers and/or receives things. We will alternate among the terms "thimac," "thing" and "machine" according to the context.

The five actions (also called stages) form the foundation for thimacs. Among the five stages, flow (a solid arrow in Fig. 7) signifies conceptual movement from one machine to another or among a machine's stages. The M's stages can be described as follows.

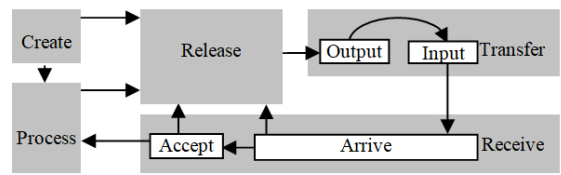

Fig. 7. The machine M

- *Arrival*: A thing reaches a new machine.
- *Acceptance*: A thing is permitted to enter the machine. If the machine always accepts arriving things, then arrival and acceptance can be combined into the "*receive*" stage. For simplicity's sake, this paper's examples presume the existence of a receive stage.
- *Processing* (change): A thing undergoes a transformation that changes it without creating a new thing.
- *Release*: A thing is marked as ready to be transferred outside of the machine.
- *Transference*: A thing is transported somewhere outside or imported inside the machine.
- *Creation*: A new thing is born (created) within a machine. A machine creates in the sense that it finds or originates a thing, brings a thing into the system and then becomes aware of it. Creation can designate "bringing into existence" within the system because what exists is what is found.

In addition, the TM model includes memory and triggering (represented as dashed arrows) that reflects relations among the processes' stages (machines).

## 4. Underpinning Software Engineering

Dynamism is this paper's underlying topic. Specifically, this dynamism is analyzed in the context of modeling communication and information in software engineering as a field that has a double nature of science and engineering, where engineering is knowledge and not just an application [18]. This section presents the main results of the paper that falls at the intersection of software engineering, conceptual modeling and the SWCM.

### 4.1 Why the Shannon–Weaver Communication Model?

According to Clark et al. [19], "Software engineering lacks underpinning scientific theories both for the software it produces and the processes by which it does so."

They propose that an approach based on information theory can provide such a benefit as research that is, among other things, based on the quantification of information involved, but more qualitative uses for information theory will be useful. Information theory is useful to software engineering as a way to consider and solve problems, and it is a contender for an underlying theory that could produce laws, explanations and predictions. Clark et al. [19] stated that information theory offers three things to software engineering:

1) A theory of communication and encoding that allows us to view software, specifications, verification and validation in a unified way as a collection of information transformation channels (information transformers) [20].

2) Due to Kolmogorov and others, a theory of objects' information content that allows us to assess the complexity of objects and to compare them at a highly useful level of abstraction [21]. This is the notion of the information in an object which Brooks asserted would be useful to software engineering [22].

3) A non-mathematical theory of information that is useful for describing the process and human aspects of software engineering, such as how individuals and teams best pool their resources to efficiently develop software systems [23].

### 4.2 Communication and Information

In the contemporary world, information and communication together occupy a dominant position as a topic in science and engineering. Models play an essential element in studying information and communication in many scientific disciplines. In this context, one difficulty is that it is often necessary to deal with heterogeneous systems from dissimilar domains. This is true with regard to modeling communication systems.

Shannon and Weaver (1949) [24] introduced the model shown in Fig. 8 (supplemented with the create action of the FM model), which led to the development of many other models. The model gives the amount of information, I, in an individual message x, and is given by: $I(x) = -\log p_x$.

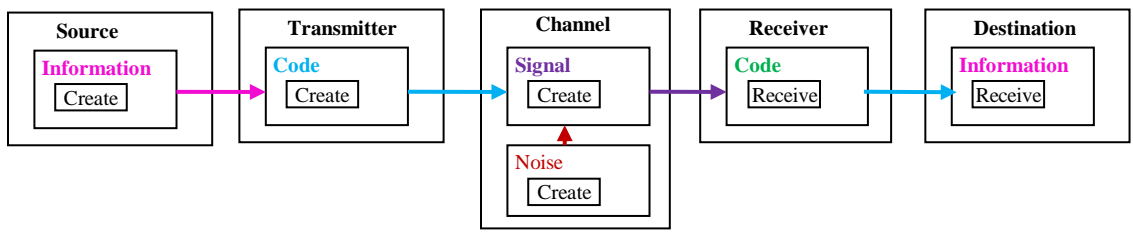

**Fig. 8. The Shannon and Weaver communication model.**

In Shannon's definition, information is proportional to uncertainty. A high level of uncertainty is associated with a high amount of information. A low level of uncertainty is associated with a low amount of information. Information is a measure of the freedom of choice in selecting a message.

> The greater this freedom of choice, the greater the information, the greater is the uncertainty that the message actually selected is some particular one. Greater freedom of choice, greater uncertainty, [and] greater information go hand in hand [24].

Shannon's definition does not cover the actual content of the messages interpreted as propositions [25]. There has been effort to formulate more semantic theories of information [26][27]. Dretske [28] analyzed the philosophical implications of Shannon's theory, "but the exact relation between various systems of logic and theory of information are still unclear" [25]. According to Firestone [29],

> Efforts to transcend Shannon's information are increasingly common…. Some of the new efforts apply the notion of statistical complexity….Others focus on the idea of 'likelihood'…. Still others develop entirely new alternative measures and theories…. It is no exaggeration to say that Information Theory is exploding today…. But, it is also true that this explosion has not yet reduced the interpretive, semantic aspect of information to a formal or physical model…. Thus the analysis and measurement of information still remains unconquered territory.

In general, the efforts to create a universal definition in which the different discipline-specific aspects would be synthesized yielded little that was new and did not lead to a universally recognized definition [3]. Kauffman [31] observed that Shannon did not tell what information is. Meaning is buried in the minimal choice. "By who or what? How? With respect to what feature of what is the choice made? What is the consequence in the real world of making the choice? How is that consequence achieved?" [31]

In fact, presently there are still many controversies about the concept of entropy (how much information), whose deep implications can be easily compared to those resulting from debates about the meaning of the term "information" [32]. Many researches have tried to clarify what "information" means. According to Floridi [27], "this is the hardest and most central problem in the philosophy of information," and that "information is still an elusive concept."

This paper focuses on the dynamic of a system modeling [24] in the "origin" (source) of the SWCM and specifically in information that is embodied in material phenomenon that produces from the data of a physical system on the left side of Fig 8. Additionally, we look at this data from the classical informational view, not as quantum phenomena. In Russell's philosophy of perception [33], molecules have no color, atoms make no noise, and electrons have no taste. Such things are verified through their relation to sense-data. Russell regarded sense-data as being part of the actual subject-matter of physics.

Fig. 9 shows an enhanced version of SWCM where the "transferrin" of things is between components: source, transmitter, receiver and destination are considered as a type of channel. The noise is removed because it does not contribute to the current level of discussion. Thus, at the start, the physical data flows to the transmitter, then the usual SWCM [24] of transmitter-channel-receiver and lastly the code is transferred to the destination. At the destination, we find the old idea that information eliminates uncertainty (entropy).

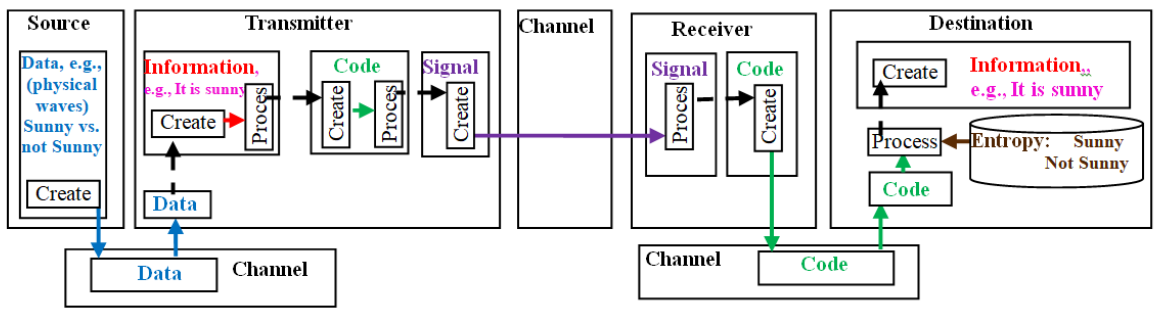

Fig. 9. An enhanced version of Shannon and Weaver communication model.

## 5. Physical Information

In this section, we apply the TM modeling to physical phenomena with a special interest in the nature of dynamism in the information creation context.

### 5.1 "Unaddressed" Information

Flückiger [30] was interested in developing a unified concept of information. According to him, we not only find diverse definitions of information, but certain fundamental problems are not treated by the different approaches at all. For example, the so-called "unaddressed" information, that is, information an individual observes by accident, is accepted by some theories as information, but explained by none of them, e.g.,

> A man is walking in a street on a very windy day. Chance would have it that a tile falls from a roof directly towards the man. If the man notices the tile, he will be informed without warning; not by binary selection from alternatives, neither by a sign, but merely by the situation of the tile falling towards him.

### 5.2 Staticity

Fig. 10 shows this scenario's TM static model of the portion of the world (the domain). This domain is tied to the diagram that represents past, present and future. It is a wholesome description that appears in reality.

Fig. 10 is an account of a physical "process." Physical processes that appear in reality are perceivable because they "radiate" *data*. Fig. 10 is assumed to be in reality and not a dream, hence, the scenario above is written as the result of perceiving it through some type of data. Data are *things* (i.e., created, processed, released, etc.) that are generated by the mere appearance (create) of a thing in nature.

In Fig. 10, the box to the left expresses that *There is a tile* (Create – circle 1). In *abstracting* this description, there is no interest in other details beside *There is a tile*. There are no details about whether this tile is on the roof of a house or a building, or whether it is red, green, etc. The tile transfers to a windy environment (2). The orientation of abstraction ignores the possibility that the tile falls to the inside of the structure it is on instead flying in the wind. The wind carries the tile as a flying (process – 3) object. The given narrative does not include the possibility of it just falling down and not flying with the wind.

The point here is that the diagrammatic picture represents the "whole" of the tile and the man narrative. The model is static in the sense that it includes all sub-scenarios consisting of the tile hitting or not hitting the man. The "whole" requires all potentialities related to the narrative. The figure also shows the source, channel and destination in the Shannon and Weaver sense where the transmitter, receiver and channel between them are eliminated.

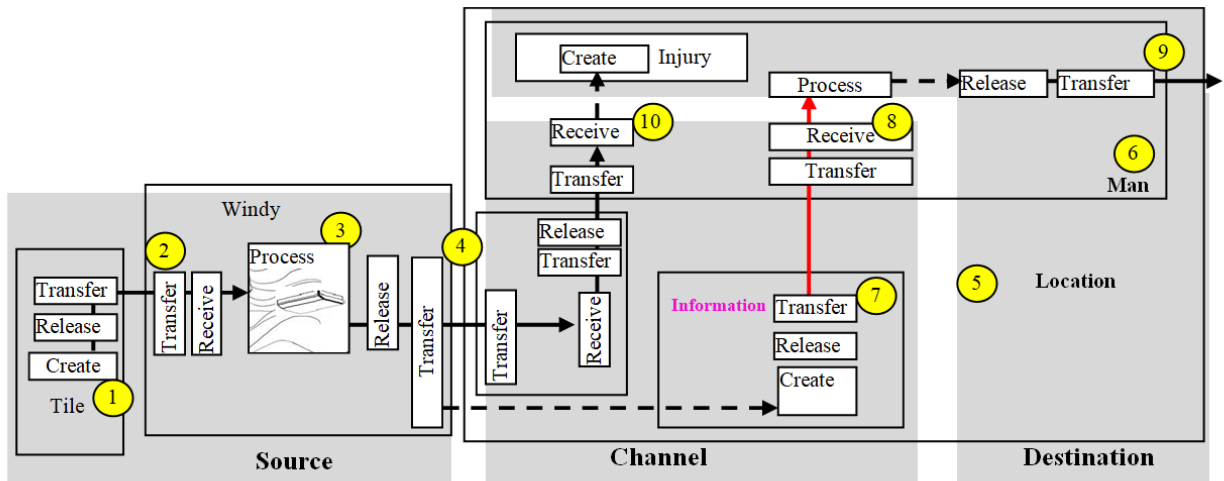

Fig. 10 The TM static model of the tile scenario.

In Fig. 10, the tile flows (4) to a certain location (5) where there is a man (6). The other possibilities are of no consequence in this narrative. For example, when we roll six-sided dice, the interest is on the top number, say 5, and there is no interest whether 3 is on the left or the right side of the dice. The flow of the tile to the location triggers creating data (7). Data, the elements of perception, are created everywhere across Fig. 10, but in this narrative, the interest is on what leads to the narrative's climax; other data are ignored if they are immaterial for this direction.

The data (7) flow to the man (8). The creation of data happens while the tile is flying in the direction of the man's position; hence, the triggering is originated while it is "transferred" in that direction. Processing such data creates information that makes the man change his location (9). If the man does not leave the location, the tile hits him and causes injury (10).

### 5.3 Dynamism

The model in Fig. 10 represents the first phase of the generalized SWCM [24] that includes source → channel → transmitter of information regarding whether the tile hits the man. Fig. 10 is a description with no consideration of time. It is a record of existence (e.g., anything that has existed in the scenario). In this subsection we examine the "happening" in this conceptual picture over time.

Dynamism refers to "change" over time. In Fig. 10, the tile does not reach instantaneously the man's position when it detaches from its original position on the building. Hence, the data signal does not reach immediately the receiver when it is transferred by the transmitter. These "changes" evolve locally with time.

Before injecting the time element into staticity, we have to decompose Fig. 10 into "regions" of events because the dynamic of a system generally is "local." For example, in the case of any human being's behavior as a system, eating generally happens "in" the hands and mouth, not in the head or toes, and walking occurs "in" the legs (not the lips), watching TV materialized "in" eyes (not the toes), etc. The dynamic aspects of a system involve the system's internal workings and originates from classical mechanics [5]. "The prototype problem was describing the motion of the planets. For this problem it was natural to give a complete characterization of the motion of all planets"[5].

The system of planets has all subsystems active. Nevertheless, there is also relative locality. The behavior of our solar system consists of a relatively stable sun with the planets rotating around the sun. This is an extreme system where most parts are continuously in the dynamic state. We can imagine a system where only one portion is active at a certain time, while the others are waiting for their dynamism turn.

To make the dynamic aspects arise in Fig. 10, we need to identify parts that are activated at different times. This converts the oneness (of the diagram) to the constellation of subdiagrams, that is, converting the diagram of Fig. 10 into subdiagrams. The composition goes with the "changes" or states:
State 1: The tile is at rest.
State 2: The tile flies with the wind.
State 3: The tile moves in the direction of the position where the man stands.
Etc.

Fig. 11 shows the decomposition of the whole diagram of Fig. 10 into parts that make "changes" or "states" that create data. These states are called events when each state is supplemented with time. For example, Fig. 12 shows the event *The data flows to the man*.

Accordingly, The events in Figure 11 are as follows.
$E_1$: There is a tile.
   − → Creating data (not shown in the figure)
$E_2$: The tile moves with the wind.
   − → Creating data (not shown in the figure)
$E_3$: The tile flies with the wind.
   − → Creating data (not shown in the figure)
$E_4$: The tile transfers *in* the area where there is a man.
− → Creating data (($E_5$)
$E_6$: The tile arrives *in* the area where there is a man.
− → Creating data (not shown in the figure)
$E_7$: The *data* about the tile transfer (see $E_4$) flows to the man.
$E_8$: The man processes the data and creates information.
$E_9$: The man changes his position.
   − → Creating data (not shown in the figure)
$E_{10}$: The tile hits the man.
   − → Creating data (not shown in the figure)

A sequence of states running in the direction of time are events that organize themselves into a combined event. Fig. 13 shows the chronology of events of this tile example with two potential outcomes $E_9$ and $E_{10}$.

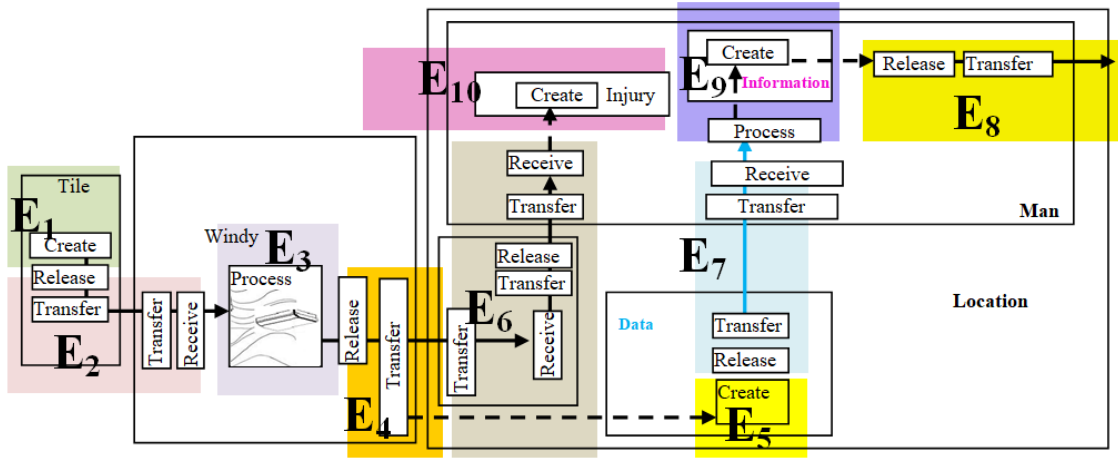

Fig. 11 Decomposition of the TM model of the tile scenario.

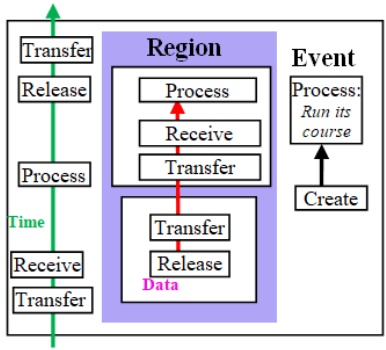

Fig. 12 The event *The data flows to the man*.

This paper claims that information in a physical phenomenon originates from data, which in turn originates in states and changes among states. The states are multiplicities that replace the wholeness of the system. These multiplicities have a logical order and time imposes further order.

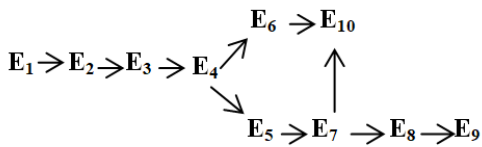

Fig. 13 The behavioral model of the tile and man domain.

## 6. Coin Tossing

The familiar coin tossing is typically used as an example of a random phenomenon. According to Nakajima [34], "the observation of reality produces phenomena, whereas the reality is confirmed (or constituted) by phenomena.

The ordinary concept of information presupposes reality as a source of information, whereas another type of information (known as it-from-bit) constitutes the reality from data (bits)." Wheeler [35] declares that every item of the physical world derives its ultimate significance from an immaterial source and explanation. According to Wheeler [35], reality arises from the posing of yes–no questions, that all things physical are information–theoretic in origin.

Consider tossing a coin (two equally likely alternatives) and its representation in the TM model. As shown in Fig. 14., we use this time the thimac representation in some parts of the model to illustrate that a thimac is a thing and a machine, as shown in Fig. 14. On the left side of the figure, we see the coin thimac (circle 1). There "exists" (i.e., created/appears) a coin as a thing (2) and as a machine (3). The coin flows in the "containing" larger thimac (e.g., "the world") where it is processed (flipped – 4), which triggers (5) the appearance (6) of a coin that is face-up (7) or triggers (8) the appearance (9) as tail-up (10). The data flows to the transmitter (11 and 12). There, the data are processed to trigger the creation of information (13), hence, the code is generated (14) and sent (15).

To specify the system's dynamism, we examine the events in this phenomenon as shown in Fig. 15. Specifically, we focus on the following events,
$E_1$: The coin appears in its initial state.
$E_2$: The coin is moved and flipped.
$E_3$: The coin lands with the face up.
$E_4$: The coin lands with the tail up.
$E_5$: The data flows to the transmitter.

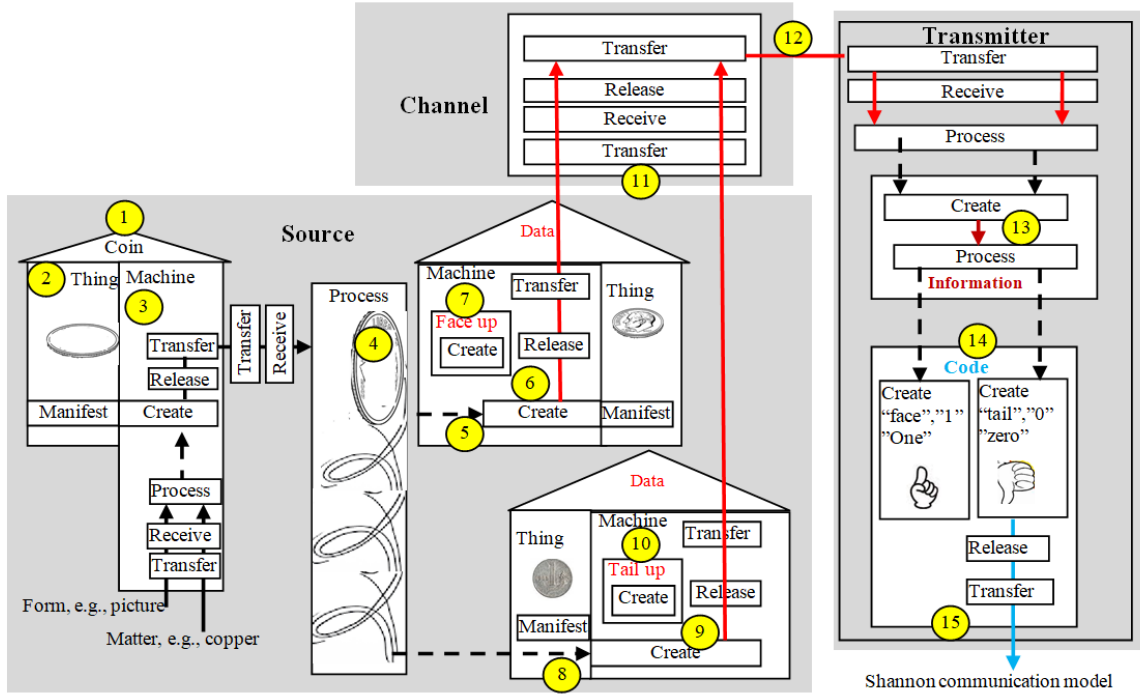

Fig. 14 The TM model of tossing the coin.

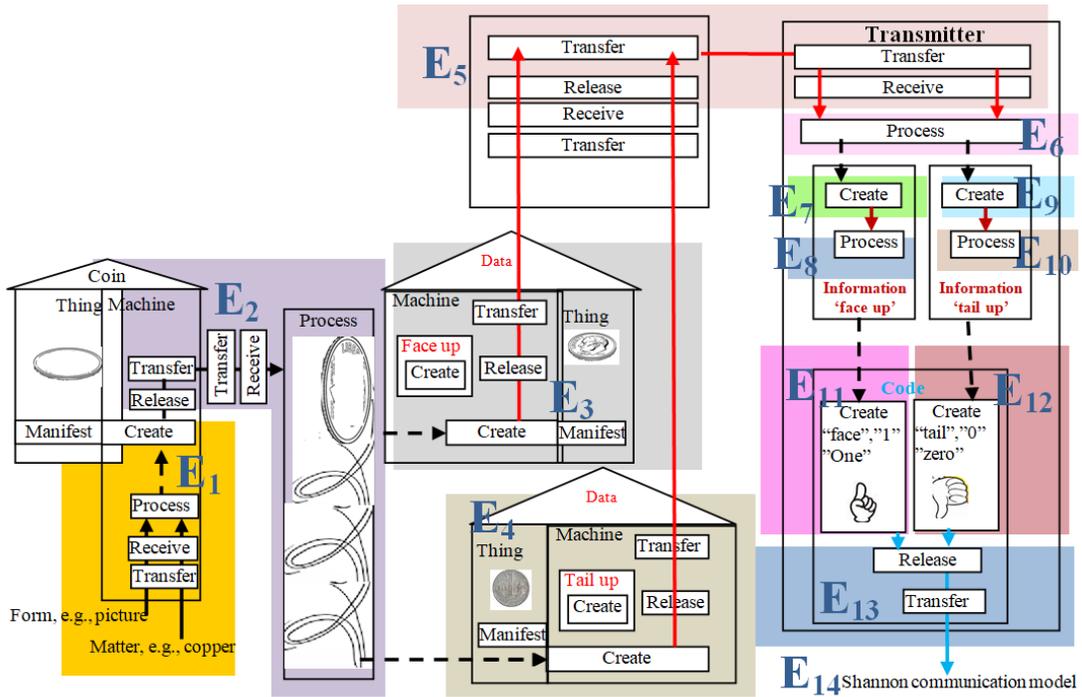

Fig. 15 The TM dynamic model of tossing the coin.

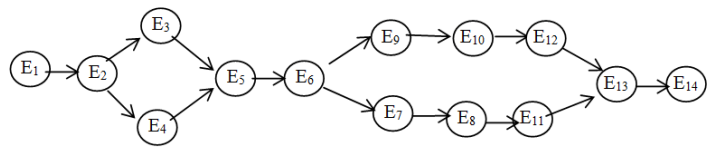

Fig. 16 The TM behavioral model of tossing the coin.

$E_6$: The transmitter processes the data.
$E_7$: The transmitter creates the information that the face is up.
$E_8$: The information that the face is up is processed.
$E_9$ The transmitter creates the information that the tail is up.
$E_{10}$: The information that the tail is up is processed.
$E_{11}$: Creating the code of "face up"
$E_{12}$: Creating the code of "tail up"
$E_{13}$: The code is sent out.
$E_{14}$: The rest of SWCM.
Fig. 16 shows the behavioral model of the coin tossing system.

The data that generates information (or the increase in information) is created in $E_3$ or $E_4$. In $E_1$, the coin manifests its "roundness," its "solidly" but not its faces. The regions of $E_3$ or $E_4$ are "impregnated" with the emergent data and they are the "seeds" of information.

## 7. Conclusion

In this paper, we first investigated the notion of dynamism in mathematical and conceptual modeling. The conclusion is that representing dynamism by equations and graphs is a form of system behavior in a conceptual modeling. Then, we focused on applying the Shannon–Weaver communication model in studying "physical" information, specifically in examining unaddressed information and in information embedded in coin tossing. Unaddressed information obeys an enhanced Shannon–Weaver communication model. Information originates in physical data that are generated by the system's dynamism. Similarly, coin tossing creates physical data that are processed to create information.

The paper introduced a modeling approach that integrates processes (dynamism) in communication in physical systems. The modeling tool seemed to provide a unified way to study different aspects of systems that leads to understanding the underlying foundation.

## References

[1] J. Edwards, "Leibniz's Aristotelian dynamism and the idea of a transition from metaphysics to corporeal nature," Substance, Force, and the Possibility of Knowledge: On Kant's Philosophy of Material Nature. University of California Press, 2000.

[2] M. Yearworth, A Brief Introduction to System Dynamics Modelling 4th Ed. University of Exeter, 27th May 2020. Accessed Sept. 9, 2020. https://www.grounded.systems/wp-content/uploads/2020/05/SD-Introduction-MY-20200527.pdf

[3] J. Koperski, Models, The Internet Encyclopedia of Philosophy, ISSN 2161-0002. https://iep.utm.edu/models/

[4] R. Frigg, and S. Hartmann, "Models in science," The Stanford Encyc. of Phil. Spring 2020. https://plato.stanford.edu/archives/spr2020/entries/models-science/.

[5] K. Aström and R. Murray, Feedback Systems: An Introduction for Scientists and Engineers, Version v2.10b. Woodstock, Oxfordshire, UK: Princeton University Press, 2009.

[6] S. Al-Fedaghi, "Modeling the realization and execution of functions and functional requirements," Int. J. Comput. Sci. Inf. Secur., vol. 18, no. 3, March 2020.

[7] S. Al-Fedaghi and D. Al-Qemlas, Modeling Network Architecture: A Cloud Case Study, International Journal of Computer Science and Network Security, vol 20, no. 3, March 2020.

[8] S. Al-Fedaghi and H. Alnasser, "Modeling network security: Case study of email system," Int. J. Adv. Comput. Sci. Appl., vol. 11, no. 3, 2020.

[9] S. Al-Fedaghi and M. Al-Saraf, "Thinging the robotic architectural structure," 2020 3rd International Conference on Mechatronics, Control and Robotics, Tokyo, Japan, Feb. 22–24, 2020.

[10] S. Al-Fedaghi, "Modeling physical/digital systems: formal event-B vs. diagrammatic thinging machine," International Journal of Computer Science and Network Security, vol. 20, no. 4, pp. 208–220, 2020.

[11] S. Al-Fedaghi and E. Haidar, "Thinging-based conceptual modeling: Case study of a tendering system," J. of Comput. Sci,

vol. 16, no. 4, 452–466, 2020. DOI: 10.3844/jcssp.2020.452.466.
[12] S. Al-Fedaghi and B. Behbehani, "How to document computer networks," J. of Comput. Sci., vol. 16, no. 6, 423–434, 2020. DOI: 10.3844/jcssp.2020.723.434.
[13] S. Al-Fedaghi, "Thinging as a way of modeling in poiesis: Applications in software engineering," Int. J. Comput. Sci. Inf. Sec., vol. 17, no. 11, November 2019.
[14] S. Al-Fedaghi and J. Al-Fadhli, "Thinging-oriented modeling of unmanned aerial vehicles," Int. J. Adv. Comput. Sci. Applic., vol. 11, no. 5, 610–619, 2020. DOI: 10.14569/IJACSA.2020.0110575.
[15] S. Al-Fedaghi and B. Behbehani, How to document computer networks," J. of Comput. Sci., vol. 16, no. 6, 423–434, 2020. DOI: 10.3844/jcssp.2020.723.434.
[16] S. Al-Fedaghi and J. Al-Fadhli, "Thinging-oriented Modeling of Unmanned Aerial Vehicles," Int. J. Adv. Comput. Sci. Applic., vol.11, no.5, 610-619, 2020. DOI: 10.14569/IJACSA.2020.0110575.
[17] S. Al-Fedaghi, Changes, States, and Events: The Thread from Staticity to Dynamism in the Conceptual Modeling of Systems, International Journal of Computer Science and Network Security, vol. 20 no. 7, pp. 138–151, July 2020.
[18] M. Lázaro, and E. Marcos, "Research in software engineering: Paradigms and methods, proceedings of the 17th International Conference," CAiSE 2005, vol. 2, Porto, Portugal: June 13–17, 2005.
[19] D. Clark, R. Feldt, S. Poulding, and S. Yoo, Information Transformation: An Underpinning Theory for Software Engineering, ICSE '15: Proceedings of the 37th International Conference on Software Engineering, Volume 2, pp. 599–602, Florence, Italy, May 2015.
[20] T. M. Cover, and J. A. Thomas, Elements of Information Theory, 2nd ed. Hoboken, NJ: John Wiley & Sons, 2012.
[21] P. M. Vitanyi, F. J. Balbach, R. L. Cilibrasi, and M. Li, "Normalized information distance," in Information theory and Statistical Learning. Springer, pp. 45–82, 2009.
[22] F. P. Brooks, Jr., "Three great challenges for half-century-old computer science," J. of the ACM, vol. 50, no. 1, pp. 25–26, 2003.
[23] K. Krippendorff, Information theory: structural models for qualitative data. Sage, vol. 62, 1986.
[24] C. E. Shannon, and W. Weaver. The Mathematical Theory of Communication. Urbana: University of Illinois Press, 1949.
[25] Adriaans, Pieter, "Information," The Stanford Encyclopedia of Philosophy, Edward N. Zalta (ed.), Spring 2019. URL = https://plato.stanford.edu/archives/spr2019/entries/information/
[26] B-H., Yehoshua, and Rudolf Carnap, "Semantic information," The British J. for the Phil. of Sci., vol. 4, no.14: 147–157, 1953. doi:10.1093/bjps/IV.14.147
[27] L. Floridi, The Philosophy of Information, Oxford: Oxford University Press, 2011. doi:10.1093/acprof:oso/9780199232383.001.0001
[28] F. Dretske, Knowledge and the Flow of Information, Cambridge, MA: The MIT Press, 1981.
[29] J. M. Firestone. Riskonomics: Reducing risk by killing your worst ideas. Prepublication excerpt, Chapter 2: What is Knowledge? 2006. http://www.kmci.org/media/Whatknowledgeis%20(non-fiction%20version).pdf
[30] F. Flückiger, Towards a unified concept of information: Presentation of a new approach, The J. of New Paradigm Research, vol. 49, issue 3–4, pp. 309–320, 1997. https://doi.org/10.1080/02604027.1997.9972637
[31] S. Kauffman. What Is Information? Cosmos and Culture, 13.7, June 4, 2010. http://www.npr.org/blogs/13.7/2010/06/04/127473541/what-is-information
[32] O. Lombardi, F. Holik, and L. Vanni, What is Shannon information? Synthese 193, 1983–2012, 2016. doi: 10.1007/s11229-015-0824-z
[33] B. Russell, Our Knowledge of the External World, 1ts ed., Philidelhia, PA: Routledge, March 6, 1993.
[34] T. Nakajima, Unification of Epistemic and Ontic Concepts of Information, Probability, and Entropy, Using Cognizers-System Model, Entropy, 21, 216; 2019. doi:10.3390/e21020216
[35] J.A. Wheeler, Information, physics, quantum: The search for links. In Proceedings of the 3rd International Symposium on Foundations of Quantum Mechanics, Tokyo, Japan, 28–31 August 1989, pp. 354–368, Cambridge, UK: Perseus Books: 1999; pp. 309–336.